\documentclass[11pt,a4paper]{article}
\usepackage{geometry}  
\usepackage[british]{babel} 
\usepackage[T1]{fontenc}
\usepackage[utf8]{inputenc}
\usepackage[pdftex]{graphicx} 
\usepackage{color}
\usepackage[hyphens]{url}
\usepackage{fourier,charter,booktabs} 
\usepackage{enumerate}
\usepackage{enumitem} 
   \setlist{labelsep=1ex, itemsep=0ex, parsep=0.2ex, topsep=0.2ex}
\usepackage{varioref}\labelformat{equation}{(#1)}
\usepackage[bookmarks=true]{hyperref}
\usepackage[round]{natbib}
\usepackage{amsfonts}

\newcommand{\Rlang}{{\normalfont\textsf{R}}{}}

\title{\bf On the use of splines for representing ordered factors}
\author{{\Large Adelchi Azzalini} \\  \large 
  Dipartimento di Scienze Statistiche \\
  Università degli Studi di Padova, Italia
 }
\date{\small \today} 

\begin{document}
\maketitle
\vspace{1ex}

\begin{quote}
\textbf{Abstract} 
In the context of regression-type statistical models, the inclusion  of
some ordered factors among the explanatory variables requires the conversion 
of qualitative levels to numeric components of the linear predictor.
The present note represent a follow-up of a methodology proposed by
\citet{azzalini:2023} for constructing numeric scores assigned to the 
factors levels.
The aim of the present supplement it to allow additional flexibility of
the mapping from ordered levels and numeric scores.

\end{quote}
\vspace{1ex}


\maketitle


\section{Including ordered factors in a linear predictor}


\subsection{Introduction}  \label{s:intro}

Consider a regression-type statistical model, such as a Generalized Linear Model 
(GLM) or alike, where one or more ordered factors must be included 
in the linear predictor. 
Denote by $F$ one such factor; in this case, the specification `ordered'
implies that there exists a natural ordering of its levels, $F_{1}, F_{2} \dots, F_{K}$.
A trite example is provided by the levels `very poor' , `poor', \dots, `excellent'
of the factor `product quality'. 
Since these qualitative terms cannot be included directly in the linear predictor
of a GLM or similar models, the question exists of constructing  appropriate 
numeric quantities for including $F$ in the linear predictor. 
This long-standing question has generated substantial debate and various 
procedures for tackling it in practice.

The present contribution, which falls within the above-delineated context,
represents a follow-up of the  proposal presented by \citet{azzalini:2023},
referred to as A2023 for the rest of the paper.  Since that paper includes an extended 
discussion  of the pertaining literature and methods, we not replicate it here, 
and only recall the essential portion of that material.
If appropriate, a reader can refer to A2023 for additional information
and discussion. 

The simplest long-established way of converting $F$ into a set of numeric
scores is simply to associate $F_{1}, F_{2} \dots, F_{K}$ to the values
$1, 2, \dots, K$. While this option is attractive for its
simplicity, it makes the questionable assumption that the effects 
of the factor levels are equally spaced. A better option, recommended
for instance by \citet{grau:korn:1987} among others,
is to select scores ``based on the substantial meaning'' of the levels,
when this is possible. While this option is definitely preferable,
it is very rarely feasible in practice; hence people often fall back
to the basic scores $1, 2, \dots, K$.

To improve on the plain use of basic scores, the currently accepted 
method of choice  involves using a set of covariates  
``coded so that individual coefficients represent orthogonal polynomials 
if the levels of the factor were actually equally spaced numeric values'',
as stated by \citet[p.\,32]{cham:hast:1993}.
Operationally, in the \Rlang\ computing environment \citep{R-lang:2024}, 
ordered factors are handled in this way unless an alternative choice is 
explicitly made; 
see the documentation of command \texttt{contr.poly} for details.
For a factor with $K$ levels, a polynomial of degree up to $K{-}1$ 
can be employed, but a smaller degree may be adopted, 
if adequate for the the data at hand.


\subsection{An alternative method}   \label{s:method-general}

An alternative to the introduction of a set of covariates in the linear predictor, 
associated to  orthogonal polynomials, has been put forward in A2023. 
We summarize it briefly, assuming for simplicity of exposition that there
is a single ordered factor $F$. The proposal involves 
the construction  of a set of numeric scores $x_{1}, x_{2}, \dots, x_{K}$ 
associated  to the levels $F_{1}, F_{2} \dots, F_{K}$.
These scores are selected by optimizing a data-fitting criterion
to be summarized later on.
From these scores, we generate a single explanatory variable $X$, according to
the available observations. 
The construction of a set of numeric scores  and of a
single column $X$ in the design matrix represent a marked difference
from the above-recalled methodology based on orthogonal polynomials,
which typically leads to multiple columns of the design matrix.

The construction of a set of scores $x_{1}, x_{2}, \dots, x_{K}$ represents 
an evolution of the basic scores $1, 2, \dots, K$, with the advantage that
their choice is driven by the data, rather than being arbitrarily set.
Consequently, their spacings may be thought of as `true' representatives
of the spacings among the levels $F_{1}, F_{2} \dots, F_{K}$.
There is then an advantage in interpretability of this logic with respect 
to other schemes, in particular with respect to the use of orthogonal polynomials.

It could be  objected that also orthogonal polynomials and associated 
regression coefficients can be used to generate numeric scores, 
but this operation involves a somewhat tortuous path, not so appealing
from an interpretation viewpoint.
  
Before describing the scheme employed for selecting the values 
$x_{1}, x_{2}, \dots, x_{K}$, some preliminary remarks are appropriate.
\begin{enumerate}[label=(\roman*)] 
\item 
It is implicit in the adoption of the qualification `ordered factor' 
that the scores are ordered, at least weakly; 
otherwise the term `ordered' would become meaningless. 
Most of what follows assumes strict inequalities in the 
ordering.
\item \label{affine}
The scores are to be determined up to an affine transformation, 
that is, the set $x_{1}, x_{2}, \dots, x_{K}$ is equivalent to 
$a{+}b\,x_{1}, a{+}b\,x_{2}, \dots, a{+}b\,x_{K}$
for any arbitrary $a$ and non-zero $b$. 
This holds because these numeric values generate the explanatory 
variable $X$  to be inserted in the linear predictor, which in turn
includes an arbitrary intercept (in most cases) and 
a free coefficient for $X$.
\item \label{increasing}
Without loss of generality, we can assume an increasing order,
so that $x_{1}< x_{2} < \dots < x_{K}$,
taking into account the previous remark.
\item 
Without loss of generality, we can assume $K\ge 3$, 
to avoid trivialities.
\end{enumerate}


Consider now a certain regression-type model (linear, GLM, or other type), 
involving a response  variable and a number of explanatory variables, 
with an associated  vector parameter $\beta$. 
The procedure for selecting the values $x_{1}, x_{2}, \dots, x_{K}$ 
is summarized as follows.
\begin{enumerate}[label=(\alph*)] 
\item 
Denote by $X$  the explanatory variable representing numerically 
the ordered factor $F$, assuming  for
the moment that the scores $x_{1}, x_{2}, \dots, x_{K}$ 
have already been selected, and consequently  $X$ is determined.
\item 
Estimation of $\beta$ is determined by optimization of a certain
target criterion,  $Q(\beta)$, say. For instance, in the case of a
GLM, $Q(\beta)$ represents the residual deviance.
\item 
Introduce a parametric function  $s(k, \theta)$ for mapping
the factor levels to real numbers
\begin{equation}
   s(k, \theta): F_{k} \to x_{k}, \qquad k\in\{1,\dots,k\}
   \label{e:s-mapping}
\end{equation}
so that, for each choice of the parameter $\theta$, a specification 
of $X$ is implied.
\item 
Then, the estimation process involves optimization of the modified
target criterion $Q_{s}(\beta, \theta)$.
Note that, when $\theta$ and hence $X$ are fixed, numerical 
optimization of $Q(\beta)$ is typically performed conveniently,
thanks to existing efficient techniques of computational statistics. 
Hence a numerical of search over the domain of $\theta$ 
involves a reasonably manageable effort, 
at least for datasets  of reasonable size.
\item
While this exposition has considered a single factor $F$,
the same scheme can be applied similarly also in the presence of 
several factors. 
The above-described procedure can simply be applied to each factor 
separately, with an increased number of components of $\theta$.
\end{enumerate}

Key to success of the procedure is the choice of the function  
$s(k, \theta)$. We want it to be a flexible function, to adapt 
to the data behaviour as closely as possible, but at the same time
we want to achieve this flexibility with a reduced dimensionality
of $\theta$. Also, it must be an increasing function, considering
condition \ref{increasing} above.

In A2023, the choice for $s(k, \theta)$ was the quantile function 
of a probability distribution regulated by parameter $\theta$, 
by interpreting $k$ as a label for the probability $k/(K+1)$.
Given remark \ref{affine} above, only the shape parameters
of the distributions actually enter in $\theta$. 
Since, usually, flexible parametric distributions feature four
parameters, two of which are for location and scale regulations,
there remain two components to regulate skewness and kurtosis;
hence $\theta$ typically comprises two components for each factor. 
Also, for numerical convenience, the focus of the development has been 
on distributions with easy-to-compute quantiles. 
Operationally, the methodology is available in the
\Rlang\ package \texttt{smof} \citep{azzalini:R-pkg-smof}.


\section{Using splines for selecting scores} \label{s:method-spline}

Numerical examples presented in A2023 and in additional instances
have indicated that the proposed methodology works satisfactorily
in most cases.
However, there was also an indication that room for further improvement
exists, by increasing flexibility of the function $s(k, \theta)$. 
The underlying reason is that parametric families of distributions, 
hence of quantile functions, are  very smooth functions, 
while there are cases where the best choice for the scores  
$x_{1}, x_{2}, \dots, x_{K}$ is not necessarily so smooth.
One could think of increasing flexibility of $s(k, \theta)$
by associating it to a mixture of probability distributions,
but the corresponding quantiles become harder to compute. 

An entirely different option for the function  $s(k, \theta)$ 
is represented by splines, typically in the form of cubic splines.
Splines allow a very high degree of flexibility, 
depending on the number of knots involved; hence they can meet
the flexibility which is our current aim.

Recall that it is possible to impose a condition of monotonicity on cubic splines; 
formulations of this type have been presented by 
\cite{frit:carl:1980} and \cite{hyman:1983}.

Since in the present problem shift and scale of the scores can
be chosen arbitrarily, we opt for fixing the conditions
\begin{equation}
  s(1,\theta)=1, \qquad s(K,\theta) = K \,.
  \label{e:spline-boundary}
\end{equation}
Now, select an arbitrary positive integer $m$ and  real values
$t_{1}, \dots, t_{m}$, $y_{1}, \dots y_{m}$ such that
\begin{equation}
  1< t_{1} < \cdots < t_{m} < K, \qquad 1 < y_{1} < \cdots < y_{m} < K  
  \label{e:spline-constraints}
\end{equation}
but are otherwise arbitrary. 
Adoption of one of the  algorithms by \cite{frit:carl:1980} and \cite{hyman:1983},
plus condition \ref{e:spline-boundary} and vector
\begin{equation}
   \theta = (t_{1}, \dots, t_{m}, y_{1}, \dots y_{m} )
   \label{e:spline-theta}
\end{equation}
identify a monotonic cubic spline $s(\cdot, \theta)$ which interpolates the points
\[
  (1, 1), ~(t_{1}, y_{1}), ~\dots, ~(t_{m}, t_{m}), ~ (K,K) \,.
\]
Finally, evaluation of $s(k, \theta)$ at $k=1,\dots, K$ provide the scores
$x_{1}, \dots, x_{K}$. 
These values can be used as  in \ref{e:s-mapping} and incorporated 
in the procedure (a)--(e) of Section~\ref{s:method-general}.

From a computational viewpoint, optimization of the target function, 
denoted $Q_{s}(\beta, \theta)$ in Section~\ref{s:method-general}, is still 
performed searching over the space of $\theta$ values, with maximization
with respect to $\beta$ for any given $\theta$.
However,  the constraints \ref{e:spline-constraints} require some care.
The adopted strategy has been to introduce working parameters free of any
constraint, which can be mapped to values of  $\theta$ satisfying
\ref{e:spline-constraints}, while optimization is performed with respect
to the working parameters.
A forthcoming version of the \Rlang\ package \texttt{smof}
will incorporate the spline variant of the methodology.


\section{A numerical illustration} 

We reconsider one of the numerical examples presented in A2023 
to illustrate the different working of the two variants of the 
methodology, using quantile and spline functions.

Specifically, we reconsider the diamond pricing data used in Section~3.2
of A2023. Again, since a fuller description of the data is provided
in that paper, we only recall the essential ingredients here. 
For a set of 537 diamonds,  their square-root transformed price is
expressed via  a linear model which includes a numerical variable
(carat) and  three ordered factors (clarity, color and cut).
Use of orthogonal polynomials to represent the ordered factors,
followed by removal of their components which are not significant,
leads to a linear model summarized as follows.

\begin{center}
\begin{tabular}{rrrrr}
  \hline
 & Estimate & Std. Error & t value & Pr($>$$|$t$|$) \\ 
  \hline
(Intercept) & 1.194 & 0.75 & 1.60 & 0.11 \\ 
  carat & 65.317 & 0.71 & 92.31 & 0.00 \\ 
  clarity.L & 24.154 & 1.58 & 15.25 & 0.00 \\ 
  clarity.Q & -11.701 & 1.24 & -9.42 & 0.00 \\ 
  clarity.C & 3.610 & 1.24 & 2.90 & 0.00 \\ 
  color.L & -13.271 & 0.96 & -13.76 & 0.00 \\ 
  color.Q & -1.907 & 0.89 & -2.13 & 0.03 \\ 
  color.C & 1.979 & 0.85 & 2.33 & 0.02 \\ 
  color\verb|^|4 & 3.369 & 0.78 & 4.30 & 0.00 \\ 
  cut.L & 1.882 & 0.85 & 2.21 & 0.03 \\ 
   \hline
 \multicolumn{5}{c}{Residual standard deviation: 6.74 on 527 degrees of freedom} \\
\end{tabular}
\end{center}

The same data have been processed using the proposed methodology.
First, a mapping function  \ref{e:s-mapping} based on a quantile function
has been employed. The adopted distribution was  Tukey's $g$-and-$h$
which features two parameters for each factor, in this case clarity and
color; hence $\theta$ has four components here.
The target function is the residual sum of squares or equivalently
the residual standard deviation, and the new summary table is as follows.

\begin{center}
\begin{tabular}{rrrrr}
  \hline
 & Estimate & Std. Error & t value & Pr($>$$|$t$|$) \\ 
  \hline
(Intercept) & 6.449 & 0.67 & 9.68 & 0.00 \\ 
  carat & 65.000 & 0.71 & 91.03 & 0.00 \\ 
  clarity.score & 8.617 & 0.47 & 18.52 & 0.00 \\ 
  color.score & -6.811 & 0.49 & -13.98 & 0.00 \\ 
  cut.L & 2.003 & 0.87 & 2.31 & 0.02 \\ 
   \hline
 \multicolumn{5}{c}{Residual standard deviation: 6.90 on 532 degrees of freedom}
\end{tabular}
\end{center}

The new table provides a compact and readable summary of the components role,
with a modest increase of the residual standard deviation.
The numeric scores assigned to the factors clarity and color are depicted 
in the next plots.

\begin{center}
\includegraphics[width=0.49\hsize]{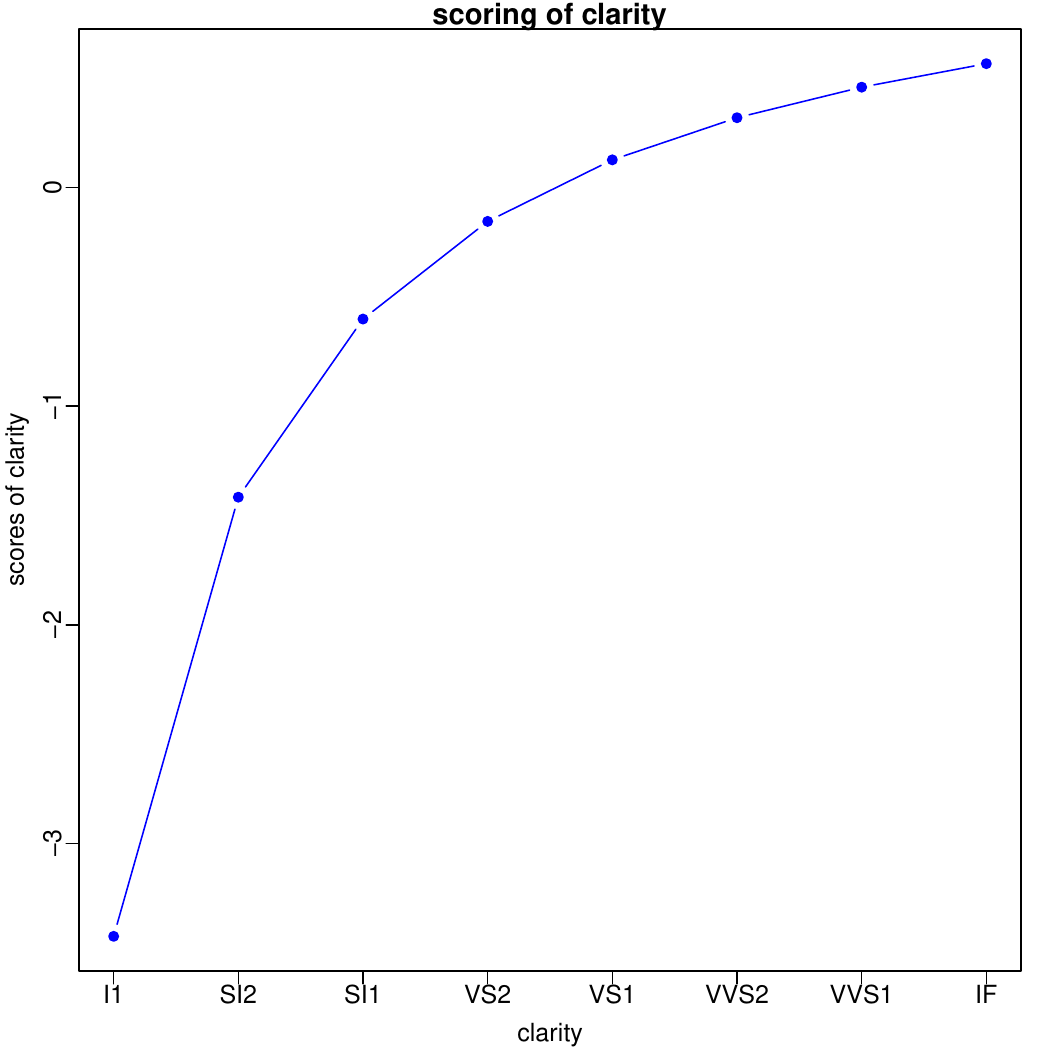}
\includegraphics[width=0.49\hsize]{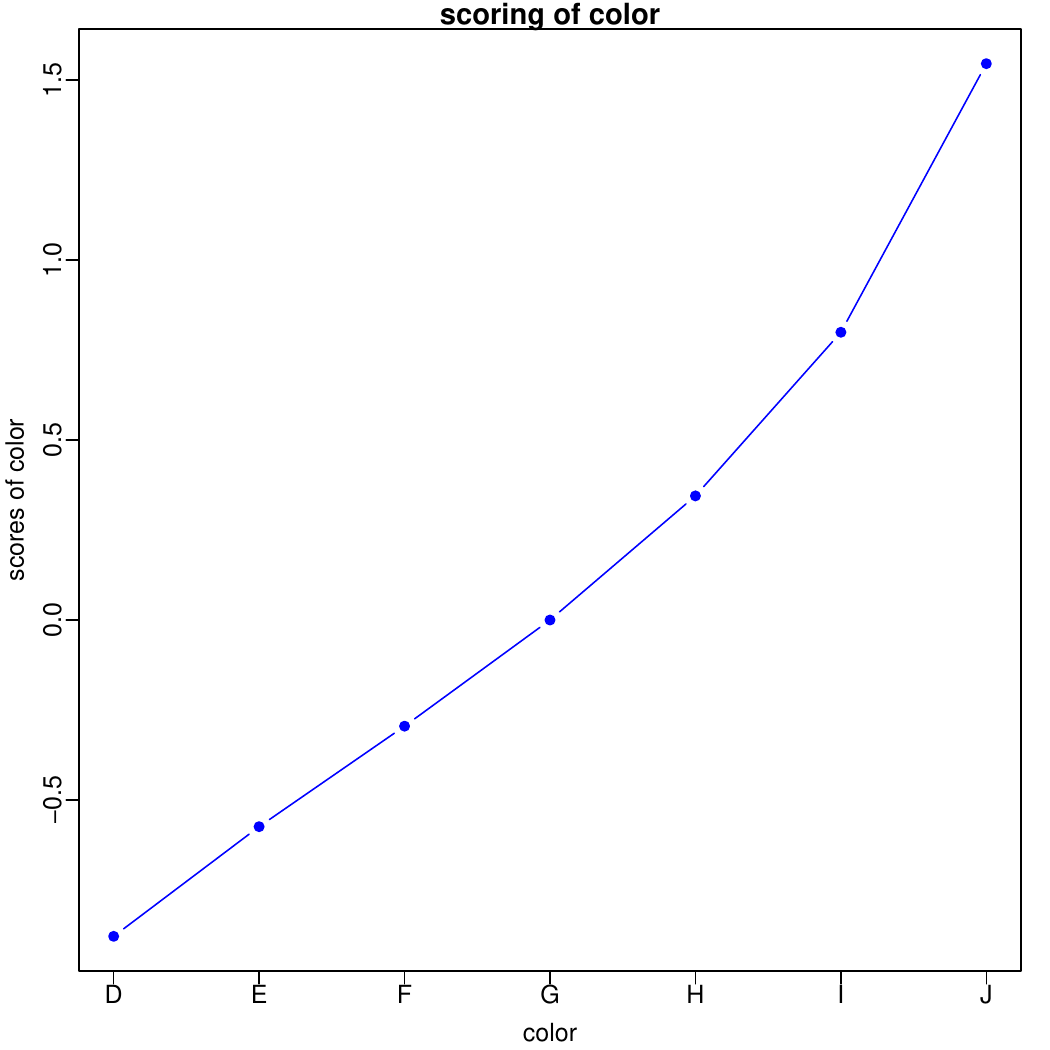}
\end{center}

In a sense, this point could be considered a satisfactory conclusion 
of the modelling process.  Suppose, however, that we are unhappy with the 
increase of residual standard deviation, although this is quite limited,
and we want to reduce it.
To increase the flexibility of the mapping function \ref{e:s-mapping},
we switch to a spline function, following the scheme described 
in Section~\ref{s:method-spline}. Specifically, the method of 
\cite{frit:carl:1980} has been employed, and  $m$ has been set to
$m=1$ for  clarity and $m=2$ for color, leading to six components
of $\theta$. The new summary table is reported next.
In the comparison with the previous table,
one should not focus on the numeric estimates of factors scores, 
since these reflect the different scales of the scores;
one should rather focus of the corresponding $t$ values 
which are largely similar to those of the earlier table.

\begin{center}
\begin{tabular}{rrrrr}
  \hline
 & Estimate & Std. Error & t value & Pr($>$$|$t$|$) \\ 
  \hline
(Intercept) & -19.317 & 1.86 & -10.40 & 0.00 \\ 
  carat & 65.238 & 0.70 & 93.34 & 0.00 \\ 
  clarity.score & 4.769 & 0.25 & 19.06 & 0.00 \\ 
  color.score & -2.242 & 0.15 & -15.17 & 0.00 \\ 
  cut.L & 1.870 & 0.85 & 2.20 & 0.03 \\ 
   \hline
 \multicolumn{5}{c}{Residual standard deviation: 6.74 on 532 degrees of freedom}  
\end{tabular}
\end{center}

It is reassuring that the residual standard deviation has been
brought back its original value, while producing simply interpretable
numeric scores  of the factors of interest.
The new plots of the factors scores indicate that factor color has 
required a more wiggly function than the one produced using the
quantile functions.  Notice that the plot does not represent the
actual spline function (although this could easily be achieved),
but shows the numeric scores evaluated at the various levels,
and the points are then joined by straight lines.

\begin{center}
\includegraphics[width=0.49\hsize]{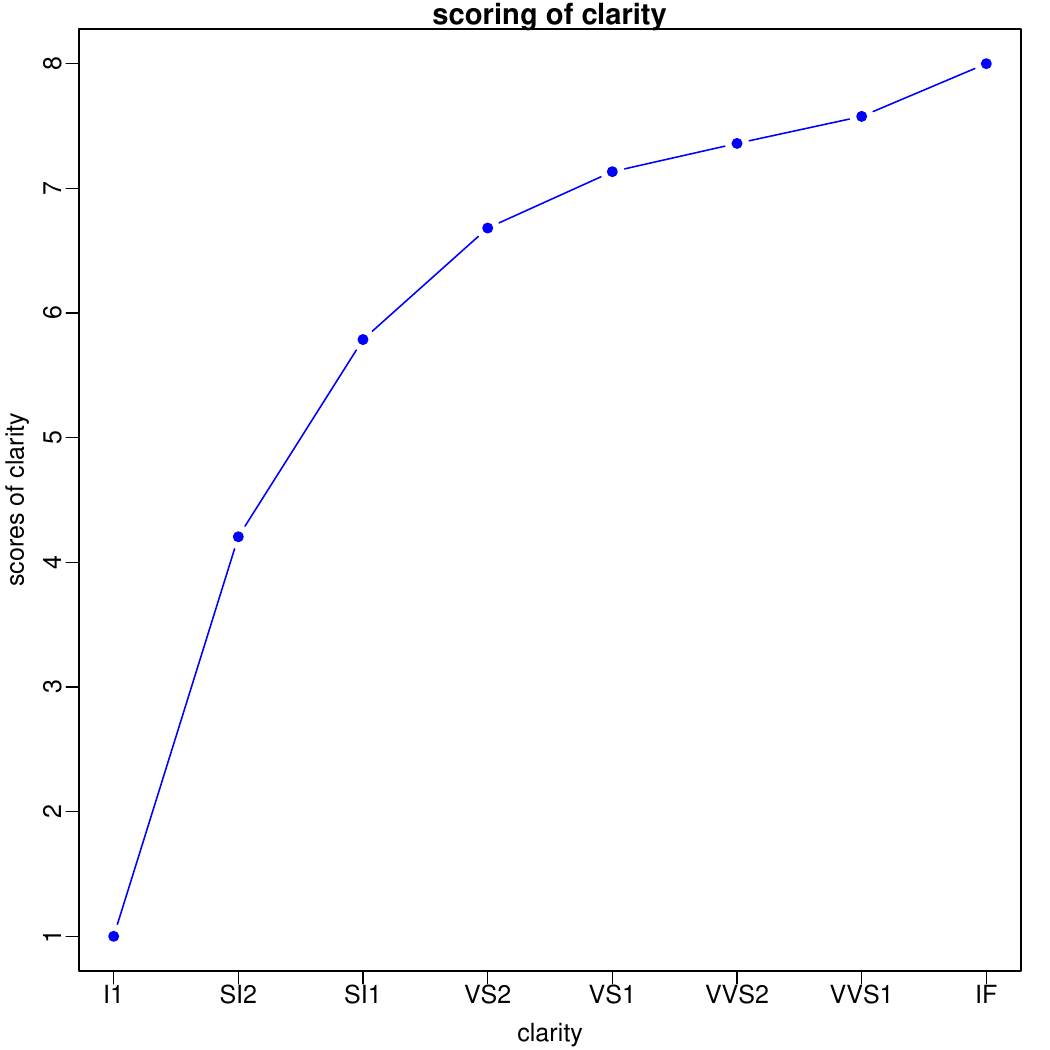}
\includegraphics[width=0.49\hsize]{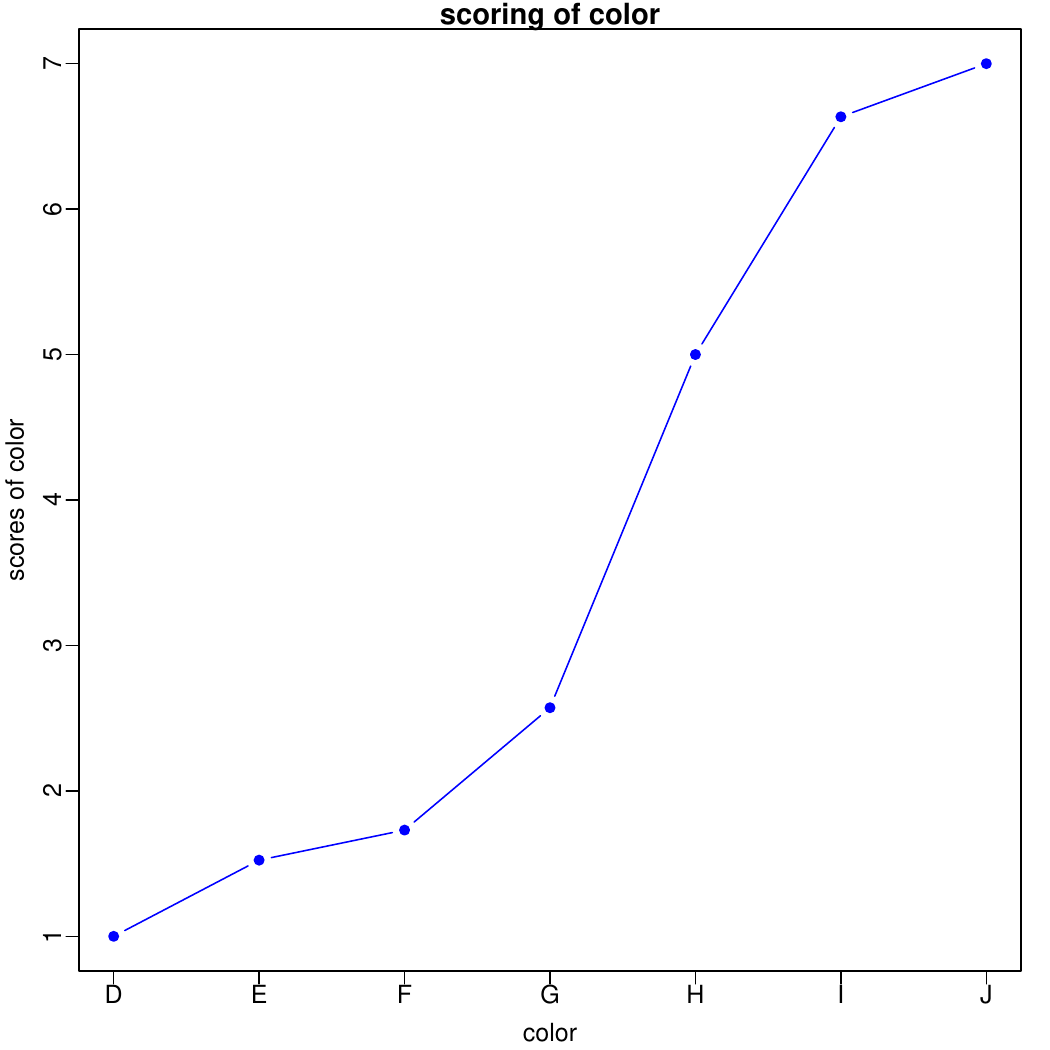}
\end{center}



\end{document}